\documentclass[conference]{IEEEtran}
\IEEEoverridecommandlockouts

\usepackage{cite}
\usepackage{amsmath,amssymb,amsfonts}
\usepackage{algorithmic}
\usepackage{graphicx}
\usepackage{textcomp}
\usepackage{xcolor}
\usepackage{graphicx} % Required for inserting images
\usepackage{pdfpages}
\usepackage{multirow, makecell} 
\usepackage{booktabs}
\usepackage{hyperref}
\usepackage{multirow}
\usepackage{multicol}
\usepackage{float}
\usepackage{amssymb}
\usepackage{csquotes}
\usepackage[figuresright]{rotating}
\usepackage{tablefootnote}
\usepackage{xcolor,colortbl}
\usepackage{threeparttable}
\usepackage{geometry}
\usepackage[many]{tcolorbox}    	% for COLORED BOXES (tikz and xcolor included)
\usepackage{setspace}               % for LINE SPACING
\usepackage{colortbl}
\usepackage{threeparttable}
\usepackage{balance}

\definecolor{main}{HTML}{5989cf}    % setting main color to be used
\definecolor{sub}{HTML}{cde4ff}     % setting sub color to be used

\newtcolorbox{boxE}{
    enhanced, % for a fancier setting,
    boxrule = 0pt, % clearing the default rule
    borderline = {0.75pt}{0pt}{main}, % outer line
    borderline = {0.75pt}{2pt}{sub} % inner line
}

\begin{document}

\title{Towards Using Personas in Requirements Engineering: What Has Been Changed Recently?}

\author{\IEEEauthorblockN{~}
\IEEEauthorblockA{\textit{~}\\
~}
\and
\IEEEauthorblockN{~}
\IEEEauthorblockA{\textit{~}\\
~}
\and
\IEEEauthorblockN{~}
\IEEEauthorblockA{\textit{~}\\
~}
}
\author{\IEEEauthorblockN{Chowdhury Shahriar Muzammel}
\IEEEauthorblockA{\textit{RMIT University, Australia}\\
s3987367@student.rmit.edu.au}
\and
\IEEEauthorblockN{Maria Spichkova}
\IEEEauthorblockA{\textit{RMIT University, Australia}\\
maria.spichkova@rmit.edu.au}
\and
\IEEEauthorblockN{James Harland}
\IEEEauthorblockA{\textit{RMIT University, Australia}\\
james.harland@rmit.edu.au}
}

\maketitle

\begin{abstract}
In requirements engineering (RE), personas are now being used to represent user expectations and needs. This systematic mapping study (SMS) aims to explore the most recent studies and to cover recent changes in trends, especially related to the recent evolution of Generative AI approaches.
Our SMS covers the period between April 2023 and April 2025. We identified 22 relevant publications and analysed persona representation, construction, validation, as well as RE activities covered by personas.  We identified that a number of studies applied AI-based solutions for persona construction and validation. We observed that template-based personas are becoming more popular nowadays. We also observed an increase in the proportion of studies covering validation aspects. 
\\
~\\
\emph{Preprint. Accepted to the 33rd IEEE International Requirements Engineering Conference Workshops (REW'25), September 1–5, 2025, Valencia, Spain. IEEE Xplore. Final version to be published by IEEE (In Press).}\\

\end{abstract}


\section{Introduction}

Requirement Engineering (RE) aims to support the development of software that fulfils the user's needs. 
One of the approaches to simplify the process of understanding user needs is to use personas.  A persona is a fictional character that represents a typical user or stakeholder, helping to understand their needs, goals, and behaviours. In the context of software design and development, the concept of personas was first introduced by Alan Cooper~\cite{cooper1999inmates}. Sim and Brouse~\cite{sim2015developing} demonstrated the application of the persona concept and ontologies, which can help to simplify the process of understanding user needs to enhance the RE process. Wang et al.~\cite{wang2025uses} demonstrated that applying a persona approach might help to increase end-user satisfaction, which is the key to successful software development.

According to Groen et al.~\cite{groen2017crowd}, Crowd-based Requirements Engineering (CrowdRE)  involves a large, extremely diversified group of users in voicing their demands and discussing their needs in the ongoing development and improvement of software products. Almaliki et al.~\cite{almaliki2015adaptive} proposed a persona-based approach to adaptive feedback acquisition, which uses personas to software feedback mechanisms for diverse users. This concept can be extended to CrowdRE to engage the crowd participants according to their feedback. 
Alamer and Alyahya~\cite{alamer2022proposed} proposed a Twitter-based strategy that helps to select the appropriate stakeholders for conducting crowdsourcing RE in app development. This user profiling can provide a basis for integrating personas in CrowdRE. Alvertis et al.~\cite{alvertis2016using} proposed the use of a tool to generate crowdsourced personas for the alignment of customer needs. The aim of the proposed tool is to generate such personas through anonymised real user profiles and data collected through third-party services. 
This demonstrates the potential of the persona approach for the interpretation of crowd-level input as human-centred artifacts that support a shared understanding and design decisions.

The latest systematic mapping study (SMS) regarding the use of personas in RE was conducted by Karolita et al.~\cite{karolita2023use} covering the time frame between January 2000 and March 2023 and providing a solid overview of research in this area. However, within the last two years, many new studies have emerged. Also, we anticipate that within this time frame, with the recent progress of GenAI, many studies regarding the use of GenAI for the persona approach might have emerged. To explore the recent studies, we conducted a replication SMS covering the period from April 2023 to April 2025. Our aim is to answer the research questions raised by Karolita et al. with respect to a more recent time frame, covering the last two years:
\\
\emph{\textbf{RQ1}  What is state-of-the-art of using personas in RE? }
\\
\emph{\textbf{RQ2} What are the benefits of using personas in RE?} 
\\
\emph{\textbf{RQ3} What are the challenges of using personas in RE?}
This would allow us to have an updated mapping that covers recent changes in the research area, especially related to GenAI, as the state-of-the-art might be significantly different to what was observed two years ago.

\textit{\textbf{Contribution:}} We conducted an SMS using a formal protocol that is based on an established guideline. The systematic search resulted in a total of 178 publications from six databases. After filtering and snowballing, we selected 22 relevant studies for analysis. To provide an updated overview of a research area, we analysed persona representation, construction, validation, and RE activities covered by personas. We also analysed how the identified studies discussed the benefits and challenges of using personas in RE. We identified that, in contrast to the research covered by the prior SMS, AI-based persona construction and validation methods have been used by a number of studies.  
We also observed that template-based persona representation is becoming more popular than in the prior period.

\section{Related Work}
\label{sec:backgroundstudy}

\subsection{Human Aspects in RE}

Grundy et al.~\cite{grundy2024ed} presented an overview of recent studies identifying key challenges of human aspects impacting software engineers and software end users. 
Human values, attitudes, behaviours, and culture can be captured using personas. Karolita et al.~\cite{karolita2023s} analysed what human factors should be covered by a persona. The researchers grouped those human aspects into three groups: `Personal characteristics', `Skill, experiential or environmental-influenced characteristics', and `Group or multiple human characteristics'. Where culture and communication style are identified as human facets within group human characteristics. 

Lachner et al.~\cite{lachner2015cross} stated that cultural persona can represent a person's behaviour, values, emotions, and expectations. They proposed a cultural persona for the Australian context to bridge the gap between theory and application. Alsanoosy et al.~\cite{alsanoosy2020cultural} also confirmed that culture can be influential in RE-related activities in their SLR. In our recent work~\cite{muzammel2024cultural}, we provided an update on this review.

Jais et al.~\cite{jais2018evidence} constructed personas to address human factors design for people living with dementia. Qyll~\cite{qyll2020persona} demonstrated that personas can incorporate human cultural attributes for decision making with the goal of having better communication and interaction in software development. 
Karolita et al.~\cite{karolita2023should} developed a comprehensive RE persona taxonomy. Based on the internal (e.g., demographic information, personality traits) and external layers (e.g., goals, motivations, pain points), the taxonomy categorizes the key human factors. Such a template can be recommended for constructing a persona in a systematic, context-specific manner.

\subsection{Personas in requirement engineering}

Droste et al.~\cite{droste2023designing} developed generalised end-user personas for explainability requirements, which are not limited to a system specifically. The authors created four personas through conducting a mixed-methods approach using an online survey and interviews. These personas will aid in eliciting general explainability requirements in early software development. So, personas can aid in understanding, communicating, and engaging diverse user communities.

Canedo et al.~\cite{canedo2023use} conducted an SLR and a survey to identify the effectiveness of personas in requirement elicitation and found that it can help to improve collaboration and communication to better understand the user needs.

Castro et al.~\cite{castro2008integrating} identified that personas can aid in the requirement analysis phase to have deep user insights, to enrich the requirement analysis process.  

A recent SMS reporting the uses of personas in RE was conducted by Karolita et al.~\cite{karolita2023use}. The authors discussed different representation methods of personas, different construction methods of personas, how they are being validated, what RE-related activities are covered by personas, the benefits of using personas, and the challenges of personas in RE in recent times. Taking this work as a background, we aim to cover our SLR regarding the recent uses of personas in RE.

\section{Research Methods}
\label{sec:methodology}

We conducted an SMS as a replication study of the SMS conducted by Karolita et al.~\cite{karolita2023use}, who claimed that they followed the guidelines proposed by Kitchenham et al. for Systematic Literature Reviews (SLRs)~\cite{kitchenham2004procedures}. 
As mentioned by Petersen et al.~\cite{PETERSEN20151}, \emph{``systematic reviews aim at synthesizing evidence, also considering the strength of evidence, systematic maps are primarily concerned with structuring a research area''}. The nature of the work conducted by Karolita et al.~\cite{karolita2023use} is more aligned with SMS rather than SLR approach, even when the corresponding guidelines~\cite{kitchenham2004procedures,PETERSEN20151} are overlapping.  
We replicated the core methodological steps of the SMS conducted by Karolita et al., but adjusted the time frame to identify the most recent studies related to the usage of personas in RE.

\textbf{Phase 1 (Automated search):} As a first step, we specified the search string. We have analysed the search string from Karolita et al.~\cite{karolita2023use} and found no need for any modifications. We will follow the same search string:

{\small{\textit{Personas AND (‘‘Requirements Engineering’’ OR ‘‘Requirements Engineering Process’’ OR ‘‘Requirements Elicitation’’ OR ‘‘Requirements Specification’’ OR ‘‘Requirements Analysis’’ OR ‘‘Requirements Gathering’’ OR ‘‘Requirements Identification’’ and ‘‘Requirements Validation’’)}}}

The prior SMS covered the time frame between January 2000 and March 2023. We specified the time frame to be from April 2023 to April 2025, and conducted an automated search in the same standard databases~\cite{karolita2023use}, except the `Engineering Village' database, which was unavailable. We decided to use 'Science Direct' instead, as both it and 'Engineering Village' access the 'Elsevier' database. 

As a result of Phase 1, we retrieved 178 papers, see Table~\ref{tab:Phase1}. % for a summary. 
We also retrieved studies from 'Science Direct' covering the original timeline (January 2000 - March 2023) and extracted 73 papers. Among them, 9 were found to be relevant, of which only 6 had already been identified in the prior SMS. We haven't included in our SMS the 3 studies that hadn't been identified by the prior SMS, as we focus on the recent trends. 

\begin{table}[h]
    \centering
    \caption{Paper retrieval information from automated search and corresponding results of selection based on IC and EC}
    \begin{footnotesize}
    \begin{tabular}{lcc}
        \hline 
        Sources & Retrieved & Selected \\ \hline \hline
        Springer & 103 & 3 \\ 
        ACM digital library & 2 & 1 \\ 
        Science Direct & 11 & 4 \\ 
        Wiley & 43 & 0 \\ 
        IEEE Xplore & 19 & 11 \\ 
        Taylor \& Francis & 0 & 0 \\ \hline
        SUM & 178  & 19 \\ \hline
    \end{tabular}
    \end{footnotesize}
    \label{tab:Phase1}
\end{table}

\textbf{Phase 2 (Filtering):} In this phase, we applied the \textit{Filtering} based on inclusion and exclusion criteria, following the Kitchenham process~\cite{kitchenham2004procedures}. After careful investigation of the criteria used in the prior SMS~\cite{karolita2023use}, we adapted those with a different time frame applied.
Thus, we applied the following \emph{inclusion criteria}:
\begin{itemize}
    \item Publications written in English only.
    \item Publications published between April 2023 and April 2025.
    \item Publications that included information about personas.
    \item Publications that focused on personas and user-centred design in RE-related tasks.
\end{itemize}
The following \emph{exclusion criteria} have been applied:
\begin{itemize}
    \item Publications that were not written in English. 
    \item Incomplete and/or short papers (less than five pages)
    \item Book chapters, prefaces, interviews, reviews, posters, panel discussions, tutorial summaries, and article summaries. 
    \item Duplicate papers (only the most complete, recent, and improved one was included if there was more than one).
    \item Papers without bibliographic information such as publication date/type, volume and issue numbers.
    \item Studies with inadequate information regarding the utilisation of personas in various RE activities. 
\end{itemize}
After applying these criteria to 178 papers identified within  Phase~1, we removed 159 publications, resulting in 19 publications to be the input for the next phase. 

%~\\
\textbf{Phase 3 (Snowballing):} In this phase, we applied \textit{snowballing} technique. We conducted forward and backward snowballing, adapting the process proposed by Wohlin~\cite{wohlin2014guidelines}. We did not get any new papers from backward snowballing. However, we got 2 more new papers by forward snowballing. We also conducted snowballing for these 2 papers, which resulted in the identification of one more relevant study. 

We also identified a number of papers that explored persona representation bias issues in LLMs, without touching any RE-related tasks. While they don't fit our inclusion criteria, this observation highlights the growth of AI-based methods for persona approaches. 
Overall, this resulted in 22 publications.

%~\\
\textbf{Phase 4 (Quality assessment):} To assess the quality of the selected papers, we applied the same quality assessment questions as specified in the prior SMS~\cite{karolita2023use}. We found all 22 identified papers of high enough quality to be included in the mapping study. 
 
%~\\
\textbf{Phase 5 (Data extraction and synthesis):} The first author first extracted data from the selected papers, then the other two authors reviewed their justification in several meetings to finalize the outcome. 

%==========================================================
\section{Results}
\label{sec:results}

In this section, we discuss what we have identified in our SMS for persona usage in RE, following the analysis by Karolita et al.~\cite{karolita2023use}. 
On this basis, we are going to answer our first research question:\\ \emph{\textbf{RQ1}  What is state-of-the-art of using personas in RE? }

We identified \textbf{22 relevant studies} published over the last two years (from April 2023 to April 2025), where 20 publications are primary studies and one paper presents an SMS conducted by Karolita et al. 
This averages 10 primary studies a year, which is a significant increase compared to the previous years. 
Karolita et al.~\cite{karolita2023use} identified 78 papers in their 22-year (January 2000-March 2023) time frame, averaging approx.  4 publications per year over 2000-2022 and approx.  5 publications per year over 2018-2022. 

Table~\ref{tab:summary} presents a summary illustrating the outcome. The summary is based on the classification applied by Karolita et al.~\cite{karolita2023use}: we analysed approaches on persona representation, construction and validation.

\begin{table*}
    
    \centering
    \caption{Overview of the persona-based studies published over April 2023 to April 2025: RE activities covered, representation, construction and validation methods.}
    \begin{footnotesize}
    \begin{tabular}{llllll}
         \hline ID &  Ref.&  Representation format&  Construction method&  Validation approach& RE activities covered\\
         \hline\hline
 
 P1 &  \cite{patkar2023data} & Template
 & Quantitative 

 & Interview

 & Validation\\
 
 P2 & \cite{nohuddin2023harnessing} &  Template
& Qualitative
 & Interview and Scenarios
 & Elicitation, validation \\

 P3 & \cite{couto2025contributions} & Template
 & Qualitative
 & Validation document
 & Elicitation
  \\
  
 P4 & \cite{otemaier2024immersive} & Template
  & N/A
 & Scenarios
 & Elicitation \\

 P5 & \cite{karolita2023should} & The prior SMS & --  & -- & -- \\

 P6 & \cite{wang2025uses} & Template \& Narrative
 & N/A
 & Interview and Survey
 & Elicitation, Specification, Validation
 \\

 P7 & \cite{gupta2023ai} & Model 
 &  AI + Qualitative 
 & AI-based
 & Elicitation, Validation
\\

 P8 & \cite{wang2023requirement} & Narrative
 & Qualitative
 & Scenarios
 & Elicitation, Analysis
 \\

 P9 & \cite{zhang2023personagen} & Template 
 &  AI + Qualitative   & AI-based
 & Elicitation, Validation
 \\

 P10 & \cite{amyot2024engineering} & Template & Qualitative & Interview 
 & Elicitation \\

 P11 & \cite{karolita2024lessons} & Template \& Narrative	& Qualitative	& Interview
 &	Elicitation, Validation
 \\
 
 P12 & \cite{droste2023designing} &  Template &	Qualitative + Qualitative 	& Validation document
 &	Elicitation
 \\

 P13 & \cite{bano2024vision} &  Visual + Model	&  
 AI + Quantitative 	& N/A
 &	Elicitation
 \\

 P14 & \cite{sureshbabu2023user} & Narrative + Visual	& Qualitative + Qualitative 	& N/A
	& Elicitation, Analysis
\\

 P15 & \cite{zulkafli2023user} & Template	& Qualitative	& N/A
	& Elicitation, Validation
  \\

 P16 & \cite{sera2024development} & Template + Visual & Quantitative	& N/A
 & 	Elicitation, Analysis
 \\

 P17 & \cite{chen2023constructing} & Template	& Quantitative	& N/A
	& Elicitation, Analysis
  \\

 P18 & \cite{kanij2023approach} & Template	& Qualitative	& Scenarios
	& Elicitation
  \\

 P19 & \cite{karolita2023should} & Template	& N/A	& Workshop	& Elicitation, Specification, Validation
  \\

 P20 & \cite{karolita2024crafter} & Template	& Qualitative	& N/A	& Elicitation, Specification, Validation
 \\

 P21 & \cite{liang2024data} & Template &	Qualitative + Qualitative 	& N/A	& Elicitation, Specification, Validation
 \\

 P22 & \cite{karolita2023s} & Narrative \& Template + Visual &	Qualitative + Qualitative	& N/A	& Elicitation, Specification, Validation
 \\
 \hline
    \end{tabular}
    \end{footnotesize}
    \label{tab:summary}
\end{table*}

%------------------------------------------------
\subsection{Persona representation}

We haven't identified any new categories of persona representation in comparison to the work of Karolita et al., where the following categories have been used:
\begin{enumerate}
    \item Text-based, which has three sub-categories:\\
      (a) Narrative form, where a storytelling approach is used; 
      (b) Template form, where the presentation should follow a particular structure or layout; 
     {(c)~Table} form, where the persona's information is presented as a table;
    \item Model-based, e.g., using UML instance diagrams, context-based story models, etc.
    \item Visual, where an approach of ``rich pictures'' or poster-like presentation is used.
\end{enumerate}
However, the overall trend of presenting personas has changed, see Table~\ref{tab:representation}. 
While in the previous SMS, the vast majority of studies used the narrative form (58 out of 78 studies, i.e. approx. 74\%), this approach has been applied only in 5 out of 21 primary studies (approx. 24\%) identified by our SMS.
Over the last two years, the majority of the primary studies used the template-based approaches (16 out of 21  studies, i.e., approx. 76\%), while as per the previous SMS, only 10 out of 78 studies (approx. 13\%) used it. We haven't identified any study using a purely table-based approach, which aligns with the results of the previous SMS: Karolita et al. identified only 2 out of 78 studies classified under this category. One of the reasons for this might be that a table-based approach has many similarities with a template-based approach (a table itself might be considered as a kind of template), so it might be reasonable to see a table as a version of a template in this context. 
In the studies we identified, the visual approach has been used in combination with other approaches, either with a narrative or with a model-based approach.

\begin{table}[ht!]
    \centering
    \caption{Persona Representation Summary}
    \begin{footnotesize}
\begin{tabular}{lllc}
\hline \multicolumn{2}{l}{Persona representation}  & Publications & Total        \\ \hline \hline
\multirow{2}{*}{Text-based} & Template  & \begin{tabular}[c]{@{}l@{}}P1, P2, P3, P4, P6, P9,\\  P10, P11, P12, P15, P16, \\ P17, P18, P19, P20, P21, P22 \end{tabular} & 16\\ \cline{3-4}
                & Narrative                           & P6, P8, P11, P14, P22 & 5 \\ \cline{3-4}
    
\multicolumn{2}{l}{Model-based}                        & P7, P13   & 2       \\ \cline{3-4}
\multicolumn{2}{l}{Visual}                            & P13, P14, P16, P22 & 4   \\ \hline
\end{tabular}
\end{footnotesize}
\label{tab:representation}
\end{table}

%---------------------------------------
\subsection{Persona construction}

The previous SMS categorised persona construction in three main techniques, following the studies introduced by~\cite{tu2010combine, jansen2021strengths, jansen2022create}:
qualitative, quantitative,
and mixed (qualitative + quantitative) techniques. 
We aim to follow the same structure. However, we identified an additional approach that hasn't been covered by the previous SMS: \textbf{AI-based persona construction}. Three out of 20 primary studies identified by our SMS used AI-based approach to generate personas:
\begin{itemize}
    \item P7~\cite{gupta2023ai} introduced AI agents that can help to generate personas with the help of end-user experiences. 
    \item P9~\cite{zhang2023personagen} used a knowledge graph and ChatGPT-4 to generate persona descriptions, which extracts user feedback and helps to create persona profiles.
    \item P13~\cite{bano2024vision} proposed the integration of AI with scenario-based prompts to generate personas for diverse users.  
\end{itemize}

\begin{table}[ht!]
    \centering
    \caption{Persona Construction Methods Summary}
    \begin{footnotesize}
\begin{tabular}{lllc}
\hline \multicolumn{2}{l}{Persona construction method} & Publications & Total \\ \hline \hline
\multicolumn{2}{l}{Qualitative} & P2, P3, P8, P10, P11 \\ & & P15, P18, P20, P22 & 9 \\ \hline
\multicolumn{2}{l}{Quantitative} & P1, P16, P17, P22 & 4 \\ \hline
\multicolumn{2}{l}{Mixed (Qualitative + Quantitative)} & P12, P14, P21, P22 & 4 \\ \hline
\multirow{2}{*}{\textbf{AI} +} & Qualitative & P7, P9 & 2 \\ \cline{2-4}
                      & Quantitative & P13 & 1 \\ \hline
\multicolumn{2}{l}{Unspecified} & P4, P6, P19 & 3 \\ \hline
\end{tabular}
\end{footnotesize}
\label{tab:construction}
\end{table}

Overall, over the last two years, only approx. 43\% of primary studies (9 out of 21) used purely qualitative methods, which is a significant difference from the previous SMS, where approx. 62\% of studies used purely qualitative methods. However, qualitative methods have also been used in combination with AI-based approaches and quantitative methods. 
Thus, the overall trend of constructing personas has changed as well, see Table~\ref{tab:construction}.

%-----------------------------------------
\subsection{Persona validation}

The set of validation methods observed over the last two years differs from the results of the previous SMS. Table~\ref{tab:validation_methods_summary} illustrates the summary of the observed persona validation methods.  
In the previous SMS, the most used method for persona validation was a focus group discussion (approx. 10\% of studies). However, none of the studies identified by our SMS used this method. The second most used method according to the previous SMS was workshop (approx. 6\% of studies), but only one of the studies identified by our SMS used workshops.

Over the last two years, the most used approach was interviews (25\% of the primary studies), while in the previous SMS only approx. 5\% of studies mentioned this method.
Other methods that have been observed in both SMSs are validation documents and scenarios. 

As the use of AI has become more prominent over the last two years, it has also impacted validation methods: two studies (10\% of the primary studies) used AI-based validation approach. This type of validation hasn't been identified in the previous SMS, and we assume that we observe an emerging and growing trend here.

Also, one of the studies identified by our SMS (P6) used a survey as a validation method, in addition to using interviews. This approach hasn't been observed previously.

\begin{table}[ht!]
    \centering
    \caption{Persona validation method summary}
    \begin{footnotesize}
    \begin{tabular}{llc}
        \hline 
        Validation Method & Paper IDs & Total Papers \\ 
        \hline \hline
        Interview & P1, P2, P6, P10, P11 & 5 \\
        \hline
        Scenarios & P2, P4, P8, P18 & 4 \\
        \hline
        Validation Document & P3, P12 & 2 \\
        \hline
        \textbf{AI-based Validation}& P7, P9 & 2 \\
        \hline
        Survey & P6 & 1 \\
        \hline
        Workshop & P19 & 1 \\
        \hline
        Unspecified & P13, P14, P15, P16, & \\ & P17, P20, P21, P22 & 8 \\
        \hline
    \end{tabular}
    \end{footnotesize}
    \label{tab:validation_methods_summary}
\end{table}

We analysed what groups of people have been involved in persona validation methods, see Table~\ref{tab:validation_people}:
\begin{itemize}
    \item Experts: Validated by domain experts via evaluation or analysis. This group of people was involved in validation in 50\% of primary studies identified by our SMS. Thus, this approach was the one most reported in the observed studies. In the previous SMS, only approx. 19\% of studies mentioned this approach out of studies presenting validation methods.  
    \item End users: Validated by real users through interviews, surveys, usability testing, etc. End users were involved in validation in 25\% of primary studies identified by our SMS. In the previous SMS, this approach was the most reported, mentioned by approx. 44\% of studies where the validation process has been discussed.
    \item Project team: Validated by practitioners through team reviews, workshops, etc. Project teams were involved in validation in 20\% of primary studies identified by our SMS. In the previous SMS, only approx. 22\% of studies mentioned this approach out of studies presenting validation methods. 

\end{itemize}
It is important to mention that out of 21 primary studies covered by our SMS, 14 studies (67\%) discussed persona validation methods and groups of people involved in the corresponding validation process.
In the previous SMS, only 35\% of studies covered this topic. This difference might highlight the growing trend in paying more attention to the validation of RE artefacts.

\begin{table}[ht!]
    \centering
    \caption{Group of people involved in  persona validation}
    \begin{footnotesize}
    \begin{tabular}{llc}
        \hline 
        Methods & Paper IDs & Total Papers \\ \hline \hline
        Experts & P3, P4, P10, P14, P15, & 10\\ & P16, P17, P18, P20, P21 & \\ 
        \hline
        End users & P2, P12, P14, P15, P21 & 5\\
        \hline
        Project team & P3, P6, P11, P20 & 4\\
        \hline
         Unspecified  & P1, P7, P8, P9, P13, P19, P22 & 7 \\ 

        \hline
    \end{tabular}
    \end{footnotesize}
    \label{tab:validation_people}
\end{table}

%-----------------------------------------------------
\subsection{RE-related activities covered by personas}

In our SMS, similarly to the work of Karolita et al., we found that personas have been used in one or multiple RE-related tasks, see Table~\ref{tab:persona_being_used}. The majority of the approaches use personas for the elicitation activities. Some approaches also propose to use personas for requirements analysis, specification and validation.

\begin{table}[ht!]
    \centering
    \caption{Persona being used in RE-related task summary}
    \begin{footnotesize}
\begin{tabular}{llll}
\hline
\begin{tabular}[c]{@{}l@{}}Nr. of task(s)\\ \end{tabular} &
  \begin{tabular}[c]{@{}l@{}}RE-related\\task(s)\end{tabular} &
  Publications &
  Total \\ \hline \hline
\multirow{2}{*}{One} 
& Elicitation & P3, P4, P10, P12, P13, P18 & 6 \\ \cline{2-4}
                     & Validation  & P1                     & 1 \\ \hline
\multirow{2}{*}{Two} &
  \begin{tabular}[c]{@{}l@{}}Elicitation,\\ Analysis\end{tabular} &
  P8, P14, P16, P17 &
  4 \\ \cline{2-4}
 &
  \begin{tabular}[c]{@{}l@{}}Elicitation,\\ Validation\end{tabular} &
  P2, P7, P9, P11, P15 &
  5 \\ \hline
Three &
  \begin{tabular}[c]{@{}l@{}}Elicitation, \\ Specification,\\ Validation\end{tabular} &
  P6, P19, P20, P21, P22 &
  5 \\ \hline
    \end{tabular}
    \end{footnotesize}
    \label{tab:persona_being_used}
\end{table}

\begin{boxE}
\textbf{Answer to RQ1:} %\\
In contrast to the prior SMS, we identified a number of studies where AI-based approach has been used for persona construction and validation. The majority of the studies used template-based representation of personas, which also demonstrates a shift in the trend, as in the prior period, the most dominant method was using a narrative form. Also, the change has been observed in the proportion of studies covering validation aspects: 67\% of primary studies covered by our SMS discussed the validation aspects, while in the prior period only 35\% of studies covered this topic. Similarly to the results of the prior SMS, we observed that the majority of the approaches use personas for the elicitation activities, which is an expected result.
\end{boxE}

%=========================================================
\section{Discussion: Benefits of using personas}
\label{sec:discussion}
In this section, we would like to summarise our observations from the studies covered by our SMS, focusing on the benefits of using personas in RE. On this basis, we are going to answer our second research question:\\
\emph{\textbf{RQ2} What are the benefits of using personas in RE?}

We identified several benefits reported in the primary studies covered by the SMS. In what follows, we discuss them in detail.

\textbf{(B1) Personas can help to represent end-users accurately}. 
This benefit has been mentioned in several studies. For example, Wang et al.~\cite{wang2025uses} concluded based on the analysis of practitioner's perspective: \textit{``...group users with different attributes, providing accurate user groups and requirements for software iterations"}. Karolita et al.~\cite{karolita2024lessons} also highlighted, based on the results of the interview-based study: \textit{``...Our participants said that personas can be a powerful method for representing users for four key reasons...''}. 
This benefit matches the results of the prior SMS, where the same advantages have been summarised under the title \emph{``Provide proper user representation.''}

\textbf{(B2) Personas can help to facilitate stakeholder communication}. Wang et al.~\cite{wang2025uses} concluded from their study, which involved interviews with 26 practitioners and a survey of 203 practitioners, that personas can serve as a proxy of humans to communicate with stakeholders.  
Also, a few participants in the study conducted by Karolita et al.~\cite{karolita2024lessons} stated that persona aids communication between stakeholders and teams. 
This benefit also matches the results of the prior SMS, where the same advantages have been summarised under the title \emph{``Communication tool between
stakeholders and developers and also between developers.''}

\textbf{(B3) Personas can be useful to evaluate design or prototype}. According to Amyot et al.~\cite{amyot2024engineering}, using of personas might be helpful in evaluating early prototypes designed for target user groups. Gupta et al.~\cite{gupta2023ai} demonstrated how to use personas directly to evaluate design artifacts. This benefit matches the results of the prior SMS.

\textbf{(B4) Personas can help to express end-users' goals and expectations}. Sureshbabu et al.~\cite{sureshbabu2023user} claimed that using personas \textit{``results in more straightforward and more natural interactions.''} Liang et al.~\cite{liang2024data} concurred, highlighting that personas ensure a clearer expression of end-users' goals and expectations. 
In the prior SMS, a similar set of advantages has been summarised as \emph{``Support better understanding of the end users.''}

\textbf{(B5) Personas can help to deal with potential human issues}. 
Weng et al.~\cite{wang2025uses} identified that using personas can help to represent potential human issues such as accessibility needs.  
Bano et al.~\cite{bano2024vision} suggested that deploying AI to generate virtual personas might help to enhance diversity and inclusion within AI systems.  
This  is partially related to the benefit \emph{``Support better understanding of the end users.''} reported the prior SMS.  However, the observed benefit B5 has a bit different focus than one mentioned in the prior SMS.

\textbf{(B6) Personas can help to identify unknown requirements}. A number of studies mentioned that personas can help to uncover requirements that weren't reported previously. 
For example, Karolita et al.~\cite{karolita2024lessons} observed that personas can uncover previously overlooked requirements. Gupta et al.~\cite{gupta2023ai} mentioned that persona can guide designers and developers in including users' needs that are generally easy to overlook. 

\textbf{(B7) Potential to streamline the use of personas through AI-based solutions}. The research conducted over the last two years by Gupta et al.~\cite{gupta2023ai}, 
Zhang et al.~\cite{zhang2023personagen}, and 
Bano et al.~\cite{bano2024vision}
demonstrated that personas can now be constructed using AI, which opens more future perspectives to use personas in RE. 

\begin{boxE}
\textbf{Answer to RQ2:} 
There are many advantages of using personas in RE. Our findings largely confirm the results of the prior SMS regarding the benefits of using personas:
(1) Personas can help to represent end-users accurately; (2) Personas can help to facilitate communication among stakeholders; (3) Personas can be useful to evaluate design or prototype; (4) Personas can help to express end-users’ goals and expectations; and 
(5) Personas can help to deal with potential human issues.
We also identified a benefit that hasn't been mentioned in the prior SMS: \emph{Personas can help to identify unknown
requirements.} We observed that \emph{AI-based approaches} can be used to construct personas and to support corresponding validation activities. This can be considered as another benefit, as it streamlines the use of personas in RE.   

\end{boxE}

%---------------------------------------
\section{Discussion: Challenges of using personas}
\label{sec:discussion2}

In this section, we would like to summarise our observations,  focusing on the challenges related to using personas. In what follows, we discuss the identified challenges in detail, 
to answer our second research question:\\ 
\emph{\textbf{RQ3} What are the challenges of using personas in RE?} 

 challenges reported in the primary studies covered by the SMS.

\textbf{ (C1) Difficulty of not dealing with end users directly}. As direct human interaction is important for communication, sometimes depending on the personas might lead to wrong requirements. 
Karolita et al.~\cite{karolita2024lessons} mentioned that one of the interviewed practitioners claimed, \textit{``The longer you physically disconnect from that person, the more your own bias comes into play again''}. 
This challenge hasn't been identified in the prior SMS.

\textbf{(C2) Validating the accuracy of personas}. Evaluating personas' accuracy is a weak point of this approach, mentioned by Liang et al.~\cite{liang2024data} and 
Karolita et al.\cite{karolita2024lessons}. 
Patkar et al.~\cite{patkar2023data} mentioned, \textit{``validating the correctness of the Personas is an open issue''}. 
Wang et al.~~\cite{wang2025uses} also highlighted that \textit{``Subjective personas without empirical data may negatively impact software development''}.
This challenge hasn't been directly mentioned in the prior SMS, but it's partially related to the challenge ``Difficulty accessing participants/data for a representable population'' identified in the prior SMS. Overall, we observed that validation aspects have become more prominent over the recent years. 

\textbf{(C3) Persona creation might include bias}. This is a common concern for personas. Multiple studies, for example, Droste et al.~\cite{droste2023designing} and Karolita et al.~\cite{karolita2024lessons}  confirmed that persona creation has the challenge of facing designer bias, which might have incorrect or inaccurate assumptions regarding actual users.
This challenge has been discussed in the prior SMS, but with a smaller scope (limited to engineers): ``Existing assumptions of engineers.'' 

\begin{boxE}
\textbf{Answer to RQ3:} %\\
The core challenges of using personas are generally related to how personas are constructed and validated. Practitioners should be aware of these limitations to be able to mitigate them. 
Our findings align with the results of the prior SMS regarding the challenges of using personas, highlighting that persona creation might include bias.
We also identified two challenges that haven't been identified by the prior SMS: (1) Difficulty of not
dealing with end users directly, and (2)~Validating the accuracy of personas. 
\end{boxE}

% %===================================================
\section{Threats to Validity}
\label{sec:threats}

We followed an established set of guidelines to conduct an SMS, but we acknowledge that our process was exposed to some threats that might affect the validity of the results.
The systematic search has been conducted in six databases that have different search string constraints. For this reason, the search strategies were varied, which might lead to missing some related studies. Also, we identified relevant papers based on the search of the titles and the abstracts. Any paper having our relevant discussion inside the paper without being explicitly mentioned in the titles and in the abstracts may be missing from our outcome. To mitigate these issues, we applied a forward and backwards snowballing strategy. 

We also acknowledge the possibility of bias while selecting the
relevant studies. To mitigate this issue, we followed the Kitchenham guidelines~\cite{kitchenham2004procedures} to specify inclusion and exclusion criteria. 
After a combined manual and automated search, we got a total of 178 papers. The first author applied both inclusion and exclusion criteria to all retrieved papers and made the primary SMS list, which is another potential threat to validity. To address this risk, the second and third authors reviewed the outcome list and proposed corrections.

\section{Conclusion}
\label{sec:conclusion}

Our study presented an updated SMS regarding the usage of personas in RE. In the last two years, there has been a significant increase in the use of personas in RE.
Our study focuses on the latest practices, benefits, and challenges related to persona-based approaches. We identified 22 relevant studies,  summarised the recent trends and core differences from the trends observed in the previous SMS. We noted the shift to template-based persona representation techniques. We analysed persona construction and validation methods and found that some approaches rely on AI-based solutions, unlike the studies identified by the prior SMS. We also observed an increase in the proportion of studies covering validation aspects.

We analysed how the identified studies discussed the benefits and challenges of using personas, and compared the trend with the previous results. Most identified benefits matched with prior SMS (e.g., improved communication), but we also identified two additional benefits: personas can help to identify unknown requirements, and there is potential to streamline the use of personas through AI-based solutions. For CrowdRE settings, this opens new opportunities to investigate how personas can identify unknown requirements automatically on a large scale.

Some of the challenges mentioned in the recent studies have been discussed in earlier studies as well (e.g., the issue that persona creation might include bias). However, two of the identified challenges haven't been mentioned by the prior SMS:  the difficulty of not dealing with end users directly and validating the accuracy of personas. Thus, it would be beneficial to develop effective persona validation techniques, especially for CrowdRE settings. 
should be used carefully, with proper validation and user feedback.

Our findings highlight that the following directions of the persona-focused RE research might be especially promising: improving AI-based automation and scalability on the one hand, and strengthening methodological processes on the other to support reliability as well as diversity and inclusion. Such a balance would be necessary for the true potential of personas to be unlocked within the practice and research of CrowdRE.

\balance 
\bibliographystyle{IEEEtran}
\bibliography{sources}

\end{document}